\documentclass[pra,twocolumn,secnumarabic,balancelastpage,amsmath,amssymb]{revtex4-2}

\usepackage{amsmath,amsfonts}
\usepackage{algorithmic}
\usepackage{mathtools}
\usepackage{graphicx}
\usepackage{textcomp}
\usepackage{xcolor}
\usepackage{physics}
\usepackage{hyperref}
\def\BibTeX{{\rm B\kern-.05em{\sc i\kern-.025em b}\kern-.08em
    T\kern-.1667em\lower.7ex\hbox{E}\kern-.125emX}}

\usepackage{enumitem}
\setlist{leftmargin=5.5mm}

\newcommand{\pL}{p_{\mathrm{L}}}

\newcommand{\pth}{p_{\mathrm{thres}}}
\newcommand{\pphy}{p_{\mathrm{phys}}}

\begin{document}
\title{Resource Analysis of Low-Overhead Transversal Architectures\\ for Reconfigurable Atom Arrays}

\author{
Hengyun~Zhou$^{1}$,
Casey~Duckering$^{1}$,
Chen~Zhao$^{1}$, 
Dolev~Bluvstein$^{2}$,
Madelyn~Cain$^{2}$, 
Aleksander~Kubica$^{3,4}$, 
Sheng-Tao~Wang$^{1}$,
Mikhail~D.~Lukin$^{2}$
}

\affiliation{
$^1$QuEra Computing Inc., 1284 Soldiers Field Road, Boston, MA, 02135, US\\
$^2$Department of Physics, Harvard University, Cambridge, Massachusetts 02138, USA\\
$^3$Department of Applied Physics, Yale University, New Haven, Connecticut 06511, USA\\
$^4$Yale Quantum Institute, Yale University, New Haven, Connecticut 06511, USA
}

\begin{abstract}
Neutral atom arrays have recently emerged as a promising platform for fault-tolerant quantum computing. Based on these advances, including dynamically-reconfigurable connectivity and fast transversal operations, we present a low-overhead architecture that supports the layout and resource estimation of large-scale fault-tolerant quantum algorithms. Utilizing recent advances in fault tolerance with transversal gate operations, this architecture achieves a run time speed-up on the order of the code distance $d$, which we find directly translates to run time improvements of large-scale quantum algorithms.
Our architecture consists of functional building blocks of key algorithmic subroutines, including magic state factories, quantum arithmetic units, and quantum look-up tables. These building blocks are implemented using efficient transversal operations, and we design space-time efficient versions of them that minimize interaction distance, thereby reducing atom move times and minimizing the volume for correlated decoding. We further propose models to estimate their logical error performance. We perform resource estimation for a large-scale implementation of Shor's factoring algorithm, one of the prototypical benchmarks for large-scale quantum algorithms, finding that 2048-bit RSA factoring can be executed with 19 million qubits in 5.6 days, for 1 ms QEC cycle times. This represents close to 50$\times$ speed-up of the run-time compared to existing estimates with similar assumptions, with no increase in space footprint.
\end{abstract}

\maketitle

\section{Introduction}
Quantum computers have the potential to be a disruptive tool for scientific discovery and practical applications~\cite{dalzell2023quantum,mcardle2020quantum,jordan2024optimization}.
Quantum error correction (QEC), where many physical qubits redundantly encode logical information and correct errors, is at the core of realizing this paradigm: It bridges the massive gap between error rates required ($<10^{-12}$)~\cite{dalzell2023quantum} and realistic physical hardware error rates ($10^{-2}- 10^{-4}$)~\cite{acharya2023suppressing,moses2023race,evered2023high}.
Unfortunately, it also comes with significant overhead, both in the number of physical qubits required to encode a logical qubit (space overhead), and in the time it takes to execute a logical operation relative to physical operations (time overhead).
Indeed, existing resource estimations for large-scale quantum algorithms, such as the execution of Shor's factoring algorithm~\cite{shor1994algorithms} on 2048-bit cryptographically-relevant instances, have ranged from 20 million qubits in 8 hours for superconducting qubits with 1 $\mu$s QEC cycle times~\cite{gidney2019how}, to several years on trapped ions and neutral atoms with 1 ms QEC cycle times~\cite{beverland2022assessing}.

Neutral atom arrays have recently emerged as a promising platform for fault-tolerant quantum computing~\cite{bluvstein2022quantum,graham2022multi}, with recent demonstrations of logical algorithms on tens of logical qubits based on a dynamically-reconfigurable architecture~\cite{bluvstein2024logical}.
A particularly powerful feature is the control-parallelism afforded by optical multiplexing: instead of controlling individual physical qubits, one can directly control an entire block of physical qubits comprising a logical qubit, allowing the implementation of transversal gate operations in a hardware-efficient way.
Unlike common gate schemes based on lattice surgery~\cite{horsman2012surface,fowler2018low,litinski2019game} or code deformation~\cite{fowler2012surface}, which require $O(d)$ syndrome extraction (SE) rounds to achieve fault tolerance due to syndrome measurement errors, transversal gates have been demonstrated to enable logical operations with only $O(1)$ SE rounds per gate for universal quantum computation~\cite{cain2024correlated,zhou2024algorithmic}.
This scheme also supports the use of fast transversal gates within magic state distillation factories, thereby allowing the high-fidelity preparation of non-Clifford resource states for universal quantum computation.
With code distances for large-scale algorithms typically in the $d\approx 30$ range, this translates into a substantial potential saving in the space-time volume of quantum computation.

With these exciting possibilities, a key milestone would be to translate these results to large-scale fault-tolerant quantum computer architectures.
There are several considerations that must be addressed.
First, the use of transversal gates influences layout, routing and timing considerations.
Second, transversal gates increase the amount of errors that each SE round must handle, requiring careful analysis of logical error rates.
Third, fast transversal gates can also lead to increases in decoding complexity if not carefully managed.
Finally, given the significant impact these factors have on the full compilation pipeline, it is important to reassess algorithmic optimizations under the new cost models and constraints.

In this work, we present a large-scale, low-overhead transversal architecture that addresses these challenges.
Our architecture framework starts from a target quantum algorithm, decomposes it into key subroutines, and then performs detailed layout and optimization of these modular gadgets.
Finally, algorithm-level parameters are re-optimized based on the resulting cost of these gadgets, in order to yield an algorithm execution that achieves the target logical error rates while minimizing space-time costs.

Our work makes the following key contributions:
\begin{itemize}
\item \textbf{Efficient transversal functional units for key algorithm sub-routines.} We introduce transversal constructions of algorithmic building blocks such as magic state factories~\cite{bravyi2005universal}, quantum arithmetic units~\cite{cuccaro2004new,gidney2018halving} and quantum lookup tables~\cite{babbush2018encoding}, designed to minimize space-time overhead, atom movement, and decoding volume, while supporting reaction-limited computation for streamlined execution.
\item \textbf{Effective models for logical performance in transversal architectures.} We introduce heuristic models for logical error rates in transversal logical circuits to capture the elevated error due to additional gate operations and the more complex decoding problem. We verify this model in key limiting cases and extract corresponding parameters.
\item \textbf{Optimization of large-scale algorithms on a transversal architecture.} With this framework, we perform an evaluation of a paradigmatic quantum algorithm as a benchmark, optimizing the implementation based on the distinct costs and considerations of a transversal implementation. We also evaluate various trade-offs and the sensitivity to parameter variation, including the effect of replacing surface code memories with dense quantum low-density parity-check (qLDPC) codes.
\item \textbf{Substantial reduction of space-time cost using transversal architectures.} With these methods, we are able to demonstrate substantial reductions in resource costs. Unlike existing estimates~\cite{gidney2019how,beverland2022assessing}, which would extrapolate to year-long computations for typical neutral atom movement and measurement times, we estimate that 2048-bit RSA factoring could be performed with 19 million qubits in 5.6 days, representing a 50$\times$ time speed-up with no increase in space footprint (Fig.~\ref{fig:comparison}). We also explain how these savings will translate to other algorithms, such as quantum chemistry. It is worth noting that this space-time saving is competitive that achieved by recent advances in qLDPC codes~\cite{bravyi2024high,xu2024constant,tremblay2022constant,higgott2024constructions}, even without considering the extra overhead for performing computation~\cite{xu2024constant,xu2024fast,stein2024architectures,cross2024improved,cohen2022low}, and the savings between the methods can likely be combined in the future.
\end{itemize}

\begin{figure}
\centering
\includegraphics[width=\linewidth]{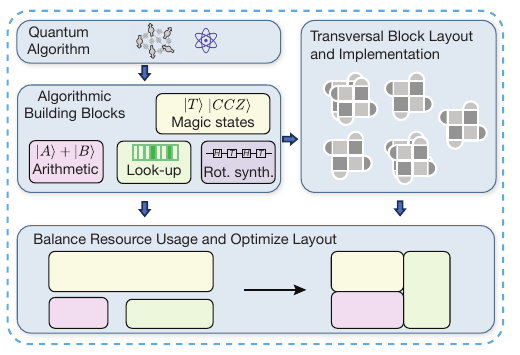}
\caption{Illustration of large-scale transversal architecture and compilation framework. A high-level quantum algorithm is decomposed into individual functional subroutines, which are implemented with transversal gates in a space-time efficient manner. Transversal gates provide an $O(d)$ logical clock speed-up, substantially reducing the run time. The subroutines are laid out and their space-time footprint is further balanced to minimize overall execution volume.}
\label{fig:overall_arch}
\end{figure}

While our framework focuses on transversal architectures for neutral atoms, we expect our results to be applicable to many other reconfigurable quantum computing platforms, such as trapped ions~\cite{pino2021demonstration,wineland1998experimental,moses2023race} and silicon spin qubits~\cite{cai2023looped}.
It also provides a common framework for which future improvements in algorithms, compilation, QEC and layout can build upon.
We therefore expect it to play a key role in guiding the construction of large-scale, fault-tolerant dynamically-reconfigurable quantum computers.

\begin{figure}
\centering
\includegraphics[width=\linewidth]{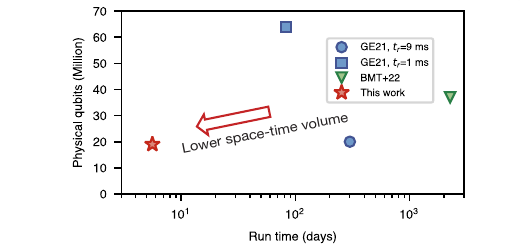}
\caption{Comparison with existing resource estimates based on lattice surgery, showing a substantial reduction in space-time volume. Data from Ref.~\cite{gidney2019how,beverland2022assessing}, where we use the attached files in Ref.~\cite{gidney2019how} to generate the blue points. See Sec.~\ref{sec:cost_breakdown} for details of the comparisons.}
\label{fig:comparison}
\end{figure}

\section{Background}
\subsection{Neutral Atom Arrays}
\label{sec:atom_array}
\begin{figure}
\centering
\includegraphics[width=\linewidth]{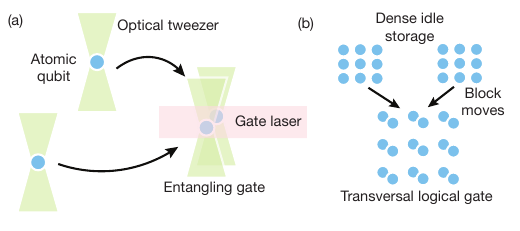}
\caption{Illustration of the neutral atom array platform. (a) Qubits are encoded in individual atoms, which can be flexibly reconfigured using optical tweezers. When two atoms are brought within the so-called blockade radius and a Rydberg laser is globally applied, they execute a two-qubit gate. (b) Qubits can be densely packed in storage, and moved in blocks. Interleaving two logical qubits and applying physical CNOTs between each corresponding pair implements a transversal logical CNOT.}
\label{fig:neutral_atom_transversal_gate}
\end{figure}

Recent advances in the neutral atom array platform make it a promising candidate for large-scale quantum computing.
In this platform, qubits are encoded in the internal states of optically-trapped atoms (Fig.~\ref{fig:neutral_atom_transversal_gate}(a)).
Long coherence times on the order of seconds to minutes~\cite{bluvstein2022quantum,barnes2022assembly,manetsch2024tweezer,jenkins2022ytterbium,ma2022universal} as well as high-fidelity single- and two-qubit gates have been demonstrated~\cite{bluvstein2022quantum,evered2023high,levine2019parallel,ma2023high,tsai2025benchmarking}.
Two key features of this platform are the dynamic reconfigurability of qubits and a high degree of control parallelism, both enabled by the natural multiplexing of optical tools such as acousto-optic deflectors (AODs).
These features open up new possibilities for exploring different fault-tolerant protocols~\cite{bluvstein2024logical}.
Atom shuttling allows long-range connectivity of qubits, enabling the use of quantum error correction codes beyond codes with 2D locality~\cite{xu2024constant,viszlai2023matching}.
Additionally, highly parallel controls facilitate the efficient implementation of transversal gates with low overhead~\cite{cain2024correlated,zhou2024algorithmic} (Fig.~\ref{fig:neutral_atom_transversal_gate}(b)).

Typical parameters for neutral atom systems are summarized in Tab.~\ref{tab:neutral_atom_params}.
The current dominant timescales are atom movement and qubit measurement.
For atom movement, the amount of time required to move a certain distance while maintaining constant thermal excitation scales as the square root of the distance, so we estimate the movement time $t$ for a given distance $L$ as
\begin{align}
t=2\sqrt{\frac{L}{a}},
\end{align}
where $a$ is the effective acceleration during the first half and effective deceleration during the second half of the trajectory\footnote{A constant jerk movement schedule is commonly employed, but the scaling of time with distance is the same. Since we estimate the acceleration from existing data on movement times and distances, the values are accurate for that schedule as well.}.

\begin{table}
\centering
\begin{tabular}{|c|c|c|}
\hline
Site spacing $l$ & Acceleration $a$  & Gate time\\\hline
12 $\mu m$    & 5500 $m/s^2$ & 1 $\mu s$\\\hline
\end{tabular}
\begin{tabular}{|c|c|}
\hline
Measure time & Decoding time\\
\hline
500 $\mu s$  & 500 $\mu s$\\
\hline
\end{tabular}
\vspace{1em}
\caption{Typical parameters for dynamically-reconfigurable neutral atom arrays, taken from Ref.~\cite{bluvstein2022quantum,bluvstein2024logical,evered2023high}. The acceleration is calculated from moving 55 $\mu m$ in 200 $\mu s$~\cite{bluvstein2022quantum}.}
\label{tab:neutral_atom_params}
\end{table}

\subsection{Quantum Error Correction}
One of the key challenges of large-scale quantum computation is noise.
In order to turn the current noisy quantum computers (physical error rate $10^{-2}$-$10^{-4}$)~\cite{preskill2018quantum} into utility-scale quantum computers with error rate $10^{-10}$ or below~\cite{dalzell2023quantum}, a logical qubit must be redundantly encoded in multiple physical qubits using QEC.
We provide an overview of the key architectural trade-offs of QEC here.
For a more detailed review see, for example, Ref.~\cite{campbell2017roads,terhal2015quantum}.

A QEC code is specified by parameters $[[n, k, d]]$, where $n$ is the number of physical data qubits per code block, $k$ is the number of error-protected \textit{logical} qubits per block, and $d$ is the code distance, which characterizes the number of detectable/correctable errors.
For many QEC codes, below a certain \textit{error threshold} $\pth$, the logical error rate decreases exponentially with the code distance, with the base of the exponent set by $\pphy/\pth$, where $\pphy$ is the characteristic physical error rate.
In order to perform QEC, one typically uses ancilla qubits to measure syndromes of stabilizers, where a value of -1 indicates the presence of an error.

One of the most important metrics for the cost of an error-corrected quantum computation is its space-time volume.
Due to the finite qubit encoding rate $k/n$, there is a \textit{space overhead} associated with the increase in qubit numbers.
At the same time, many standard QEC constructions require multiple syndrome extraction (SE) rounds to gain confidence about the values of noisy syndrome values, resulting in a \textit{time overhead} order of the code distance $d$~\cite{dennis2002topological,fowler2012surface}.
Many methods for trading off space against time exist~\cite{fowler2012time,beverland2022assessing}: for example, we can use the methods summarized later in Sec.~\ref{sec:bridge_qubit}, or simply repeat the algorithm several times in separate copies over space.
Therefore, the space-time volume of the computation, defined as the product of the qubit number and run time, is often used as the optimization objective for resource estimations~\cite{gidney2019how,beverland2022assessing}.
We will also adapt this as our objective function in our analysis below.

Finally, a critical component of a quantum error correction system is the \textit{decoder}~\cite{das2020scalable,barber2025real,liyanage2023scalable,higgott2025sparse,wu2023fusion}.
The decoder's job is to consider physical measurements results from the quantum computer execution, determine from these what errors likely occurred, and apply suitable corrections (often in software) to interpret logical measurements and determine future quantum operations.
The round-trip time from measurement to decoder, and to the next conditional quantum operation is called the \textit{reaction time} of the system~\cite{gidney2019flexible} and critically influences the speed of the quantum computer.

\subsection{Surface Codes}
The majority of our analysis focuses on the rotated surface code~\cite{litinski2019game,fowler2018low}, due to its high error threshold (around 1\% depending on the exact implementation)~\cite{fowler2012surface}, its simple structure, its support for transversal gate operations~\cite{shor1996fault,dennis2002topological}, and the ability to quickly and accurately decode and correct errors~\cite{higgott2025sparse,wu2023fusion}.

\begin{figure}
\centering
\includegraphics[width=\linewidth]{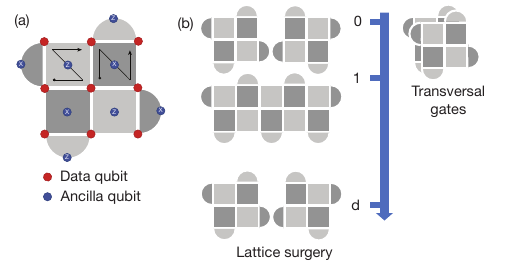}
\caption{Illustration of the surface code and various logical entangling gate schemes. (a) Atom layout of an SE round and measure qubit moves. (b) Lattice surgery is performed by joining two code patches along the edge, and requires $O(d)$ SE rounds to achieve fault tolerance. In contrast, transversal gates can be fault-tolerantly executed with $O(1)$ SE rounds per gate, by interleaving the qubits via atom movement.}
\label{fig:surface_code}
\end{figure}

One code block of the $[[d^2,1,d]]$ surface code consists of $n=d^2$ physical qubits encoding $k=1$ logical qubits, arranged in a $d\times d$ square grid of atoms.
Syndrome extraction can be performed in a distance-preserving way with a single ancilla qubit per stabilizer (Fig.~\ref{fig:surface_code}(a)).
Thus, in addition to data qubits, each surface code patch typically requires $d^2-1$ ancilla qubits.

There are multiple methods to perform logical operations in the surface code, based on code deformation~\cite{fowler2012surface,vuillot2019code}, lattice surgery~\cite{horsman2012surface}, or transversal gates~\cite{dennis2002topological} (Fig.~\ref{fig:surface_code}(b)).
The first two methods usually require $d$ SE rounds, so we focus on transversal gate operations, which can support 1 SE round per logical operation~\cite{cain2024correlated,zhou2024algorithmic}.
Transversal gates with the surface code have recently been experimentally implemented in neutral atom quantum processors~\cite{bluvstein2024logical}.

\subsection{Transversal Gates}
Transversal gates are an approach to perform logical operations on QEC codes by implementing independent physical gates on disjoint partitions of the system.
For example, a transversal logical CNOT between two code patches is performed by physical CNOTs between each matching qubit of the two codes (Fig.~\ref{fig:surface_code}(b)).
Because errors do not spread within a code block, transversal gates are inherently fault-tolerant.
Logical $X$, $Y$ and $Z$ gates are ``free'' and are performed simply by tracking Pauli frames in software.
$H$ and $S$ gates are permutation-transversal and fold-transversal respectively~\cite{fowler2012surface,moussa2016transversal,kubica2015unfolding,gidney2024inplace}, and can be implemented via a single physical gate layer followed by qubit rearrangement.
The Eastin-Knill theorem rules out a single quantum error correction code with a transversal universal gate set~\cite{eastin2009restrictions}.
However, similar to standard lattice surgery constructions, we can prepare non-Clifford resource states (magic states) using methods such as magic state distillation (MSD)~\cite{bravyi2005universal}.
We can then perform universal quantum computation using these resource states.

By performing correlated decoding across multiple logical qubits, recent theoretical work has shown that $O(1)$ SE rounds per logical operation suffices for transversal gate fault tolerance~\cite{cain2024correlated,zhou2024algorithmic}.
One of the key challenges to this approach is the increase in classical decoding complexity arising from correlated decoding, as multiple logical qubits must be decoded together.
In particular, logical qubits that are within a distance $d$ in the quantum circuit must be jointly decoded in a windowed decoding approach~\cite{dennis2002topological,skoric2023parallel,tan2023scalable,bombin2023modular}.
Therefore, designing the circuit to limit the relevant decoding volume is of key importance.
We achieve this below by judicious design of our subroutine layouts.

\section{Transversal Architecture for Large-Scale Quantum Computation}
\subsection{Overall Architecture}

Our compilation framework (Fig.~\ref{fig:overall_arch}) starts from a target quantum algorithm and decomposes it into key algorithmic building blocks.
Each of these building blocks are then further compiled into surface code transversal gate operations and explicit layouts, optimized for space-time efficiency.
In the following sections, we provide detailed descriptions of several commonly used algorithmic subroutines, including high-fidelity magic state preparation, quantum adders, and quantum look-up tables, which form the foundation of many modern quantum algorithms~\cite{gidney2019how,babbush2018encoding,kivlichan2020improved}.
These subroutine generators take as input certain parameters, such as instance size, code distance and syndrome extraction frequency, and output the layout, together with an estimate of the space and time cost of the subroutine, as well as the resulting logical error rate.
The high-level quantum algorithm may also have a variety of parameters and algorithmic choices that can be made, leading to parameterized gadget inputs and parameterized unit costs.
For example, one could choose an approach that has higher T-gate parallelism and decreased T-depth, which may lead to faster execution of the main computation, but requires a larger space footprint to support higher magic state factory throughput.

Our framework shares many common features and philosophies with other state-of-the-art large-scale algorithm compilations, such as those in Ref.~\cite{gidney2019how,litinski2022active}.
We utilize bridge qubits and time-optimal quantum computation~\cite{fowler2012time} to flexibly lay out the computation in space-time and minimize latency between non-Clifford gates with sequential dependencies (see Sec.~\ref{sec:bridge_qubit} for details).
We also make use of many of the key algorithmic innovations that have led to reduced costs, such as windowed arithmetic~\cite{gidney2019windowed} (using look-up tables and adders to more efficiently implement arithmetic operations, see Sec.~\ref{sec:qrom}, Sec.~\ref{sec:adder}) and oblivious carry runways~\cite{gidney2019approximate}.

At the same time, there are also key distinctions in our approach.
First, we employ direct atom movement for qubit routing, which provides a high degree of flexibility and helps avoid certain forms of congestion caused by the limitations of planar lattice surgery routing~\cite{beverland2021surface}.
Second, although atom movement provides flexibility and generally has a move time that scales quite favorably as the distance increases (Sec.~\ref{sec:atom_array}), long distance moves still take more time, potentially slowing down the logical clock speed if used extensively.
We therefore design layouts that keep the moves local most of the time, ensuring that the logical clock speed remains reasonable.
Moreover, we explicitly calculate the amount of time a given execution step will take based on the movement distances, in order to obtain an accurate estimate of the algorithm run times.
Third, while transversal gates enable $O(1)$ SE rounds per logical operation, they require the solution of a potentially more challenging decoding problem~\cite{cain2024correlated,zhou2024algorithmic,sahay2024error}.
To manage the complexity of this, we exploit the locality of our gadget designs, such that for many dominant gadgets, there is only a linear increase in decoding volume.
Recent advances in decoder design can further lower the cost of decoding~\cite{wu2024hypergraph,wan2024iterative,sahay2024error,cain2025fast,serra-peralta2025decoding}.
All these considerations mean that many key algorithmic parameters and choices need to be re-examined and re-optimized, which we carry out within our framework.

\subsection{Example: Factoring}
\label{sec:factoring}
We use Shor's algorithm~\cite{shor1994algorithms} as a showcase for our transversal architecture and the spatial layout of algorithm subroutines.
Shor's algorithm requires all our key algorithmic building blocks and makes for clear comparison to previous architectures~\cite{gidney2019how,beverland2022assessing,whitney2009fault}.

Given some input $N=p\times q$, the purpose of Shor's factoring algorithm is to determine the prime factors $p$ and $q$.
For $N$, an $n$-bit number, no known classical algorithm can solve this in polynomial time, whereas Shor's algorithm achieves $O(n^3)$ gate complexity.
To do so, Shor's algorithm uses quantum phase estimation (QPE) and the quantum Fourier transform (QFT) to find the period of the function $f(x)=g^x ~(\mathrm{mod}~ N)$, which can then be used to determine $p$ and $q$ via classical post-processing.
We use the Ekera-Hastad variant of the algorithm~\cite{ekeraa2020on,ekera2017}, which has the benefit of reducing the exponent length required and achieving close to unity classical processing success rate.

\begin{figure}
\centering
\includegraphics[width=\linewidth]{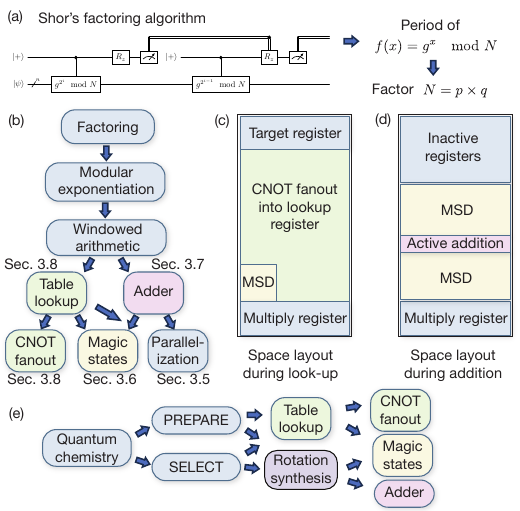}
\caption{(a) Shor's algorithm performs period finding on the modular exponentiation function. Using iterative quantum phase estimation~\cite{kitaev1995quantum}, the bits in quantum phase estimation can be processed in small groups, shown here for 1 bit per group. (b) Decomposition of Shor's algorithm into functional building blocks. (c,d) High-level layout and schematic of relative resource usage during quantum table look-up and quantum addition. (e) Decomposition of quantum chemistry problems based on tensor hypercontraction~\cite{lee2021even} into functional building blocks.}
\label{fig:shor_blocks}
\end{figure}

Using iterative quantum phase estimation (Fig.~\ref{fig:shor_blocks}(a)), the QFT portion of the algorithm only requires a single rotation per bit, and thus the cost of Shor's algorithm is dominated by the modular exponentiation in evaluating $f(x)$, where the base is a constant and the power is a quantum superposition.
This can be decomposed into controlled multipliers, and further into controlled adders.
Instead of performing these primitives directly, we employ the windowed arithmetic optimization from Ref.~\cite{gidney2019windowed}, which computes the coefficients of groups of bits classically (the group size is defined by the exponent window $w_{\mathrm{exp}}$ and multiplication window size $w_{\mathrm{mul}}$), loads them into a quantum register via a quantum look-up table (QROM)~\cite{babbush2018encoding}, and adds them into the target register.
By off-loading part of the computation onto a classical processor, this reduces the number of non-Clifford gates required on the quantum computer, which typically plays a large role in determining the resource costs.

The main cost of implementing Shor's algorithm thus comes down to performing windowed arithmetic for modular exponentiation, which we decompose into our individual subroutine generators as in Fig.~\ref{fig:shor_blocks}(b).
We follow the standard procedure of estimating the cost and number of calls for each subroutine~\cite{gidney2019how,litinski2022active}, which has been shown to be a good approximation for the overall resource cost of classical and quantum algorithms.
We lay out the computation~\cite{gidney2019how} schematically for the lookup and addition stages in Fig.~\ref{fig:shor_blocks}(c,d), where both movement within modules and the interfaces (input-output) between different gadgets are performed with local moves.

\subsection{Example: Quantum Chemistry}

While our evaluations focus on factoring as a concrete algorithm to perform extensive analysis of trade-offs and sensitivity, we emphasize that the same approach is readily transferable to other applications.
Here, we briefly comment on how a similar approach applies to quantum chemistry ground state energy estimation applications.

State-of-the-art quantum algorithms for quantum chemistry~\cite{babbush2018encoding,von2021quantum,lee2021even,harrigan2024expressing} employ a qubitization approach~\cite{low2016hamiltonian}, together with improved representations of the Hamiltonian.
The algorithm involves repeated blocks of PREPARE, which prepares a superposition $|\mathcal{L}\rangle=\sum_{l=0}^{L-1}\sqrt{\frac{w_l}{\lambda}}|l\rangle$, and SELECT, which applies the $l$th Hamiltonian term $H_l$ when the control register $|\mathcal{L}\rangle$ is in state $|l\rangle$, thereby allowing one to implement the Hamiltonian $H=\sum_l w_lH_l$.

Optimized quantum circuits for implementing the PREPARE and SELECT subroutines are described in Ref.~\cite{lee2021even,caesura2025faster}.
For PREPARE and PREPARE$^\dagger$, the algorithm cost is dominated by table lookup (see Sec.~\ref{sec:qrom}), which accounts for 90-95\% of the T-count based on the equations and Fig.~11 in Ref.~\cite{lee2021even}.
For SELECT, the algorithm cost comes primarily from rotations controlled by the $|\mathcal{L}\rangle$ register, which as shown in Fig.~5 of Ref.~\cite{lee2021even} are implemented with a table lookup (30\% of the T count) and controlled rotations (70\% of the T count), where the state-of-the-art way to implement such rotations is to perform addition with a phase-gradient state~\cite{gidney2018halving}, thus reducing the implementation of the SELECT subroutine into table lookup (see Sec.~\ref{sec:qrom}) and adders (see Sec.~\ref{sec:adder}).

Therefore, while our resource estimations focus on factoring, many other applications~\cite{gilyen2018quantum,martyn2021grand} such as quantum chemistry utilize the same building blocks as described here, and we expect the same reduced space-time volume and improved resource estimates.
We leave a detailed layout and resource estimate of such applications to future work.

\subsection{Logical Error Model with Transversal Gates}
\label{sec:scaling}

Accurate estimates of logical error rates are important for estimating the resource requirements, as the amount of error suppression determines the code distance choice, which in turn dictate the space-time cost of performing a logical operation.
Our goal in this section is to develop a heuristic formula, backed by simulation results, to estimate the logical error rate for each transversal gate.

We focus on the standard circuit-level noise model~\cite{gidney2021fault,higgott2023improved1}, in which each operation is followed by (or, in the case of measurement, preceded by) a depolarizing channel acting on each qubit.
For simplicity, we assume a uniform error rate $\pphy$, although our discussions apply more generally.

For the standard surface code memory~\cite{dennis2002topological,fowler2012surface}, sufficiently below the threshold, the logical error rate per SE round per logical qubit is exponentially suppressed with the code distance $d$, i.e.
\begin{align}
\pL = \frac{C}{\Lambda^{\frac{d+1}{2}}},\quad \Lambda=\frac{\pth}{\pphy}.
\end{align}
The parameter $\Lambda$ characterizes the rate of improvement in logical error rate as the code distance increments by 2~\cite{acharya2023suppressing,chen2021exponential}, and the prefactor $C$ has been estimated to be approximately 0.1 for the surface code~\cite{beverland2022assessing,gidney2019how}.

Intuitively, the process of QEC is a battle between the amount of errors added due to gate operations and entropy removal due to syndrome measurements.
We thus consider the amount of errors added due to both transversal gates and syndrome extraction operations themselves, for each SE round.
While there may be some variation across different logical circuits because of the varying error propagation, we assume that the contribution of each transversal gate is fixed, as suggested by numerical evidence in Ref.~\cite{cain2024correlated,zhou2024algorithmic}.
Specifically, Fig.~S3 of Ref.~\cite{cain2024correlated} shows that the logical error rate for different realizations of random logical circuits behave similarly, and the logical error rate for circuits of different depth scales proportional to the circuit depth.
Given these results, we approximate the logical error rate per qubit per SE round as
\begin{equation}
p_L = C\left( \frac{\sum_j\beta_j p_j}{\pth} \right) ^{\frac{d+1}{2}},
\end{equation}
and assume that the logical error rates are additive between different qubits and SE rounds.
Here, $p_j$ represent different error sources during syndrome extraction and transversal gates and $\beta_j$ are corresponding weights.
As the addition of transversal logical gates can lead to a more complex decoding problem, we expect that they will have a slightly larger weight $\beta_j$ compared to gates from syndrome extraction, and the value will depend on the accuracy of the decoder.

For concreteness, let us focus on a deep Clifford circuit consisting of CNOT gates, with $x$ transversal CNOTs per SE round.
We define the decoding factor $\alpha=\beta_{\mathrm{CNOT}}p_{\mathrm{CNOT}}/(\sum_{j\in SE}\beta_j p_j)$ (the summation runs over gates in the syndrome extraction only).
Intuitively, the decoding factor succinctly captures how much transversal gates increase the effective noise rate that error correction must handle.
As discussed above, with a more challenging decoding problem, $\beta_{\mathrm{CNOT}}$ and hence $\alpha$ will increase, leading to a higher effective error rate.
The value of $\alpha$ can be estimated from numerical simulations, as we show below.
Using $\alpha$, we can express an ansatz for the per-CNOT logical error rate as
\begin{align}
p_{L,\mathrm{CNOT}}=\frac{2C}{x}\qty(\frac{\alpha x+1}{\Lambda})^{\frac{d+1}{2}},
\label{eq:ler-per-cnot}
\end{align}
where the factor of 2 is because a CNOT involves two qubits, the division by $x$ is due to the cost being amortized across $x$ CNOTs, while the multiplication by $x$ is due to the elevated noise.
This expression has the important feature that when the number of SE rounds per CNOT is large ($x\rightarrow 0$), it reproduces the usual memory limit, as is desired.

\begin{figure}
\centering
\includegraphics[width=\linewidth]{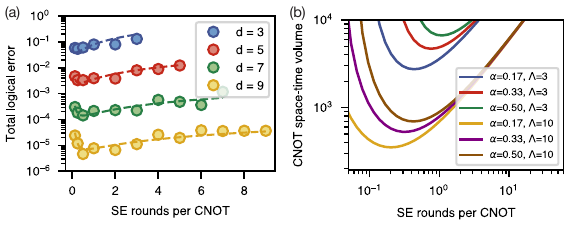}
\caption{(a) Logical error model with transversal gates and fitting with heuristic formula. We fit the depth-32 random Clifford circuit data from Ref.~\cite{cain2024correlated} to the model in Eq.~(\ref{eq:ler-per-cnot}), finding good agreement and justifying the logical error model. (b) Estimate of the space-time cost per logical CNOT based on the heuristic formula.}
\label{fig:fitting}
\end{figure}

In Fig.~\ref{fig:fitting}(a), we fit the model in Eq.~(\ref{eq:ler-per-cnot}) to the numerical data with a most-likely-error (MLE) decoder from Ref.~\cite{cain2024correlated}, for a physical error rate of $0.1\%$, finding good agreement.
This justifies the use of our heuristic model.
We find that this MLE decoder results in faster logical error suppression than the usual formula ($\Lambda$ is closer to 20 for a physical error rate of $0.1\%$, as opposed to 10 as commonly assumed).
For consistency with existing literature, we therefore use the standard values $C=0.1$, $\pphy=0.1\%$, $\pth=1\%$ and $\Lambda=10$, but use the relative coefficients obtained from the fit to characterize how much a transversal CNOT contributes, finding $\alpha\approx 1/6$.
We assume the same $\alpha$ value and code distance~\cite{gidney2024inplace} for transversal $S$ gates that mix $X$ and $Z$ bases.
For the same physical error rate, there are more operations per SE round.
Since there are more chances for errors, this results in a reduction of the threshold
\begin{align}
p_{\mathrm{thres,eff}}=\frac{\pth}{\alpha x+1},
\end{align}
which is consistent with the $\gtrsim 0.87\%$ threshold for 1 CNOT per SE round found in Ref.~\cite{cain2024correlated}.

To heuristically estimate the optimal number of CNOTs per SE round, we estimate the space-time volume per logical CNOT for different numbers of SE rounds per CNOT.
For each SE frequency, we calculate the code distance required to reach a target logical CNOT error rate per qubit of around $p_{\mathrm{targ}}=10^{-12}$, resulting in a space-time volume
\begin{align}
V\propto d^2\qty(\frac{4}{x}+1)\propto \frac{\log^2\qty(\frac{x\cdot p_{\mathrm{targ}}}{2C})}{\log^2\qty((\alpha x+1)\Lambda)}\qty(\frac{4}{x}+1).
\end{align}
Here, the first term in the parenthesis is the number of syndrome extraction CNOT gates per transversal logical gate (each SE round has 4 CNOTs), and the second term corresponds to the transversal CNOT gate.
We show the results in Fig.~\ref{fig:fitting}(b), where the optimal number of SE rounds is typically less than or equal to 1, and varies based on the physical error rate as well as decoding factor.
We also study the impact of different values of $\alpha$ at the architecture level in Fig.~\ref{fig:error_rate_sensitivity}(b), where we find that a larger decoding factor, as may be expected for other decoders such as BP-OSD~\cite{roffe2020decoding,panteleev2019degenerate}, BP-LSD~\cite{hillmann2024localized}, hypergraph union find~\cite{delfosse2022toward,cain2024correlated}, or various adaptations of matching decoders to logical algorithms~\cite{sahay2024error,wan2024iterative,cain2025fast,serra-peralta2025decoding}, only causes a small increase in the space-time volume for physical error rates of $0.1\%$.

\subsection{Parallelizing Execution Using Space-Time Trade-offs}
\label{sec:bridge_qubit}

A key method in optimizing the execution of quantum circuits is performing space-time trade-offs (time-optimal quantum computation~\cite{fowler2012time,gidney2019flexible} or bridge qubits~\cite{litinski2022active}), which allows one to optimize the ``active volume" of the algorithm execution~\cite{litinski2022active,gidney2019how}.

\begin{figure}
\centering
\includegraphics[width=\linewidth]{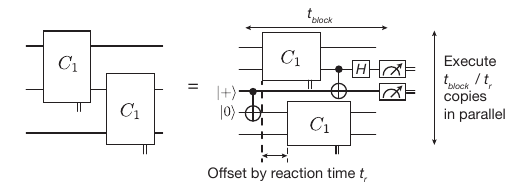}
\caption{Using Bell pairs to parallelize multiple circuit segments. Two sequentially-repeated circuit segments, as appears in the adder and table look-up gadgets below, can be executed in parallel by using a Bell pair. Due to sequential dependencies arising from non-Clifford gates, each block is offset by the reaction time $t_r$, and $t_{\mathrm{block}}/t_r$ blocks are executed in parallel. The dependent-measurements are resolved one-by-one after all relevant prior information is available, including the Bell basis measurement outcomes.}
\label{fig:bridge}
\end{figure}

This is illustrated in Fig.~\ref{fig:bridge}.
A Bell state preparation and Bell basis measurement \textit{bends a qubit backward then forward in time}, allowing the sequential blocks to be executed in parallel.
The Bell measurement results inform Z and X corrections to inject in the second block.
This allows any Clifford circuit to be implemented in constant depth.
However, implementing non-Clifford gates often requires logical measurements whose basis depend sequentially on each other, which requires a time equal to the \textit{reaction time} $t_r$ defined above to resolve.
Therefore, when executing multiple circuit segments in parallel, we offset each subsequent block by $t_r$, so that the measurement basis is determined just in time.
We make use of auto-corrected magic states~\cite{gidney2019flexible,litinski2019game}, allowing one to perform the correct non-Clifford gates simply by adapting the logical measurement basis.
With each block taking duration $t_{\mathrm{block}}$, we can execute $t_{\mathrm{block}}/t_r$ blocks in parallel.
Because not all qubits are active during the full block, in practice, we compute the number of copies required based on how long each qubit is active within the circuit.

\subsection{Key Algorithmic Subroutine: Magic State Factory}
\label{sec:msd}
We now describe our strategy to prepare non-Clifford magic states, which allow us to complete universality via magic state teleportation.
The most common strategy to prepare high-fidelity magic states from low-fidelity physical operations is to perform magic state distillation~\cite{bravyi2005universal}, in which noisy magic states are refined into high-quality magic states via error-corrected Clifford operations.
Magic state distillation with logical qubits has been demonstrated in recent experiments on neutral atom hardware~\cite{sales2024experimental}.

Recently, new strategies for preparing magic states have been proposed~\cite{jones2016gauge,choi2023fault,gavriel2022transversal,chamberland2020very,hirano2024leveraging}, culminating in the $|T\rangle$ magic state cultivation scheme~\cite{gidney2024magic} that achieves competitive overheads.
While further improvements are likely possible, the fidelities achieved currently fall just short of what is required for many large-scale applications for $\pphy=10^{-3}$.
In light of this, and considering that transversal implementations of magic state distillation are relatively cheap due to fast transversal gates, we choose to use an 8$T$-to-$CCZ$ factory~\cite{jones2013low,gidney2019efficient}, which converts 8 $|T\rangle$ magic states into a single $|CCZ\rangle$ magic state.

\begin{figure}
\centering
\includegraphics[width=\linewidth]{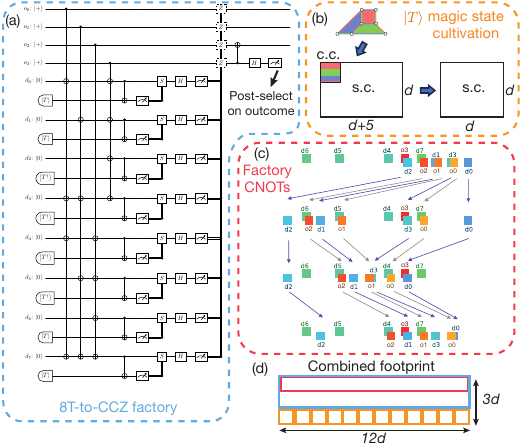}
\caption{Illustration of the magic state factory. (a) The second stage converts 8$|T\rangle$ states into a higher-quality $|CCZ\rangle$ state upon error detection.
(b) The first stage uses $|T\rangle$ magic state cultivation, which starts from a color code (c.c.) with in-place transversal Clifford gates, and grows into a larger surface code (s.c.). We extend the surface code region and measure out the color code region, so that the resulting code is a regular surface code.
(c) Movement layout for the CNOT portion of the $8T$-to-$CCZ$ factory, corresponding to the circuit in (a).
(d) Estimated total footprint of the factory, with colors corresponding to the highlighted gadgets in (a-c).}
\label{fig:factory}
\end{figure}

The quantum circuit for the 8$T$-to-$CCZ$ factory is shown in Fig.~\ref{fig:factory}(a), which makes use of the transversal non-Clifford gate of the [[8,3,2]] code~\cite{jones2013low,campbell2016,wang2024fault,bluvstein2024logical,honciuc2024implementing}.
It consists of 4 CNOT layers to prepare an entangled state between the output logical qubits and a set of logical qubits further encoded into an [[8,3,2]] code (factory CNOTs), followed by teleported $T$ gates on the [[8,3,2]] code (teleportation).
The $\ket{0}$/$\ket{+}$ input states for the factory logical qubits are prepared by initializing all constituent physical qubits in $\ket{0}$/$\ket{+}$ and performing a single SE round.
This was shown to be sufficient in Ref.~\cite{zhou2024algorithmic} when correlated decoding is applied to the logical algorithm~\cite{cain2024correlated}.
Upon measuring the bottom logical qubits, applying feed-forward Pauli corrections and post-selecting on the correct factory outcome, the output state will be 
\begin{align}
|CCZ\rangle=CCZ|+\rangle^{\otimes 3}
\end{align}
with logical error probability suppressed as
\begin{align}
p_{\mathrm{out}}=28p_{\mathrm{in}}^2+o(p_{\mathrm{in}}^2),
\end{align}
assuming that the Clifford operations in the factory are ideal due to the inner surface code protection.
We design the layout of the factory CNOT step using the optimal layout synthesis tool OLSQ-DPQA~\cite{tan2024compiling}, finding a 1D layout that does not require any qubit re-ordering~\cite{xu2024constant} (Fig.~\ref{fig:factory}(c)), thereby simplifying the control requirements for this key subroutine.
While the factory CNOT step is executing, we simultaneously perform SE rounds on the $|T\rangle$ states to grow them from the code distance of the first stage to the full distance of the factory.

Since the $8T$-to-$CCZ$ factory only provides quadratic suppression, we prepare high-quality $|T\rangle$ input states using the magic state cultivation scheme from Ref.~\cite{gidney2024magic}.
The output state of this procedure is in a ``grafted" code that contains color code and surface code regions (Fig.~\ref{fig:factory}(b) bottom left), so we extend the size of the patch during the growth stage to $d+5$-by-$d$ and measure out the strip of width 5 containing the color code, in order to prepare a regular $d$-by-$d$ surface code.
The cultivation scheme allows a continuous trade-off between $|T\rangle$ state quality and post-selection overhead, so we choose parameters such that the output fidelity of our combined magic state factory is sufficiently good for the target application.
As an example, the factoring application requires $3\times 10^9$ $CCZ$ gates; assuming that the $CCZ$ error budget should not exceed $5\%$, we target a per $|CCZ\rangle$ error of $1.6\times 10^{-11}$, which translates to a per $|T\rangle$ cultivation error of $7.7\times 10^{-7}$.
Reading from Fig.~1 of Ref.~\cite{gidney2024magic}, this yields an expected volume of $1.5\times 10^4$ qubit rounds, which we estimate is sufficient to fit 8 copies of cultivation in the $12d$-by-$1d$ bottom row of the factor in Fig.~\ref{fig:factory}.

\subsection{Key Algorithmic Subroutine: Quantum Arithmetic}
\label{sec:adder}

A quantum arithmetic circuit is equivalent to classical reversible logic, but made out of quantum gates and executed on a quantum superposition state.
Extensive work has gone into optimizing reversible modular addition, multiplication, and exponentiation circuits for use on quantum computers~\cite{gidney2018halving,cuccaro2004new,gidney2019windowed,zalka1998fast,van2005fast,draper2004logarithmic}.
Due to the use of windowed arithmetic~\cite{gidney2019windowed}, we focus on adders, which transform registers $\ket{a}\ket{b}\rightarrow \ket{a}\ket{a+b}$.

\begin{figure}
\centering
\includegraphics[width=\linewidth]{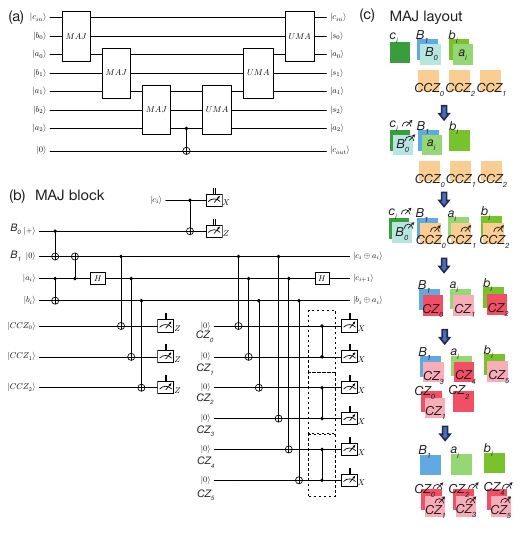}
\caption{(a) Illustration of the Cuccaro ripple-carry adder gadget~\cite{cuccaro2004new}, consisting of MAJority (MAJ) blocks and UnMajority $\&$ Add (UMA) blocks.
(b) Detailed circuit for the MAJ block, explicitly including the implementation of Toffoli gates via CCZ ancillae and auto-corrected CZ gates (bottom right block and dashed boxes, which indicate gates conditional on earlier measurement results).
(c) Detailed spatial layout of the MAJ block, which can be performed primarily in a $3\times 2$ region.}
\label{fig:cuccaro}
\end{figure}

We choose the Cuccaro ripple carry adder~\cite{cuccaro2004new} shown in Fig.~\ref{fig:cuccaro}(a) due to its low $T$ count, low requirement on space and more steady consumption of magic states compared to other methods~\cite{gidney2018halving}.
It consists of repeated MAJ and UMA blocks, where the MAJ subcircuit uses CNOTs and Toffoli gates to compute the next carry bit $c_{i+1}$ in-place, which is the majority-vote of the three input (qu)bits.
This is output on the third qubit, with the first two qubits preserving reversibility / unitarity.
After propagating all the carry bits, UMA (UnMajority and Add) undoes the majority to restore the bit-values of $a_i$ and calculates the sum bit $s_i$.

We implement the MAJ and UMA blocks with the circuit shown in Fig. \ref{fig:cuccaro}(b), using bridge qubits ($B_0$, $B_1$) to parallelize different blocks as in Sec.~\ref{sec:bridge_qubit}.
MAJ and UMA each use one Toffoli gate, which we implement using the distilled $\ket{CCZ}$ state from Sec.~\ref{sec:msd}.
We perform Clifford corrections using extra ancilla qubits ($CZ$ qubits), where the $\ket{CCZ}$ measurements determine which, if any, of the three conditional CZ gates (shown in dashed boxes) are executed.
This allows the computation to proceed in a fashion limited by the reaction time~\cite{gidney2019flexible}, as described in Sec.~\ref{sec:bridge_qubit}.
The explicit spatial layout and moves of the MAJ block are shown in Fig. \ref{fig:cuccaro}(c), where we find that the gadget fits in a $3\times2$ logical block and the maximal move distance in any step is $\sqrt{2}dl$, with $d$ being code distance and $l$ being atom site separation.
The linear ripple carry structure leads to a larger $T$-depth that ties up idle quantum resources, so we use the oblivious carry runway method to break the addition into multiple segments~\cite{gidney2019approximate}, with different segments separated by the runway separation length $r_{\mathrm{sep}}$ and padded by a runway padding $r_{\mathrm{pad}}$ to ensure the errors are sufficiently small.
The number of segments directly determines the number of $T$ factories and extra space resources needed to keep the computation running at a reaction-limited rate.
We therefore choose the number of segments such that the space requirements match that of the lookup phase described below.

\subsection{Key Algorithmic Subroutine: Quantum Look-Up Table}
\label{sec:qrom}

\begin{figure*}
\centering
\includegraphics[width=\linewidth]{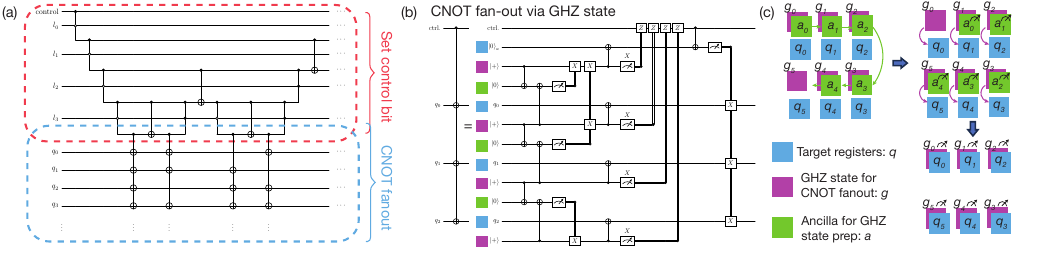}
\caption{Illustration of look-up table gadget.
(a) Overall circuit of a quantum look-up table controlled by 4 qubits and targeting a number of other qubits. The top half sets the control bit, which then performs a CNOT fan-out to load the table values.
(b) Implementation of CNOT fan-out via a GHZ state. The green ancillae measure $ZZ$ on an all $|+\rangle$ initial state, thereby performing measurement-based preparation of the purple qubits in a GHZ state. A transversal CNOT to the target registers then performs the desired CNOT fan-out.
(c) Spatial layout of the CNOT fan-out, showing that local movement suffices.}
\label{fig:lookup}
\end{figure*}

Quantum look-up tables (also known as QROMs~\cite{babbush2018encoding}) allow loading classical data into quantum registers in superposition.
As the name QROM (Quantum Read-Only Memory) suggests, given the address $\ket{l}$ on the top register of qubits, a QROM will load some pre-defined, classically-calculated bits into the lower register $\ket{q}$ (Fig.~\ref{fig:lookup}(a)).
The additional control qubit enables or disables the operation of the entire circuit.
A look-up table with $m$ control qubits in $\ket{l}$ can contain up to $2^m$ entries, so we only use it with $m=O(\log n)$-sized windows to look up small blocks of coefficients for modular exponentiation.

The look-up table circuit efficiently loops through all bit values of $l$ using temporary AND gates (Toffoli gates acting on $|0\rangle$)~\cite{gidney2018halving}, setting the bottom qubit in the red block of Fig.~\ref{fig:lookup}(a) as 1 at the $i$th step when the control qubit is in state $\ket{l=i}$.
This qubit then updates the target qubits, using CNOT gate fan-outs (single solid control connected to multiple NOT targets based on the table entry).

The top half of the circuit has a similar structure to the adder circuit, and can be implemented in a similar fashion.
Each lookup table entry requires 1 Toffoli and 1 CNOT on average, which is a much smaller cost compared to the CNOT fan-out for register sizes relevant for factoring.

The CNOT fan-out ostensibly requires long-range interactions, so we must ensure long atom-move distances and growth of the decoding volume do not become a bottleneck.
For example, a naive implementation might use a log-depth circuit to achieve the required fan-out, necessitating long moves~\cite{low2024trading}.
We instead create GHZ resource states ($\frac{1}{\sqrt2}\left(\ket{0\dots 0}+\ket{1\dots 1}\right)$) using measurement-based methods~\cite{nielsen2010quantum}, and use this to assist the CNOT fan-out~\cite{fowler2018low,kim2022fault,pham2012d,baumer2024measurement}.
As shown in Fig.~\ref{fig:lookup}(b), by preparing GHZ qubits in $|+\rangle$ (purple squares), performing two layers of CNOT gates to ancilla helpers (green squares) prepared in $|0\rangle$, and measuring the helper qubits, a GHZ state can be prepared.
Transversal CNOTs from the GHZ state to target registers, followed by measurement of the GHZ qubits in the $X$ basis, allows one to perform the desired operations.
In Fig.~\ref{fig:lookup}(c), we identify a layout that places all qubits close to their target locations, with the ancillas snaking through the 2D plane, thereby ensuring that only moves of a small, constant distance ($2dl$ in this illustration) are required.
Typical entries in the table will have around half of the values active, so in principle, we only need half of the qubits in the GHZ state.
We strike a balance between qubit usage and move distance by using an underlying GHZ state grid of a certain spacing (this parameter is optimized over in our experiments), and then add additional qubits based on the target register locations as needed.
Because there are multiple stages of the GHZ state preparation and consumption, there are opportunities for pipelining, which we also consider and optimize over in our evaluations.
We find that typically, having a single copy in each stage of the pipeline minimizes the total space-time volume for the parameters we consider.

\section{Evaluation}
Having described our overall framework and architecture for transversal implementations of large-scale quantum algorithms, we now carry out a detailed analysis of the resource costs of such an implementation, and evaluate various available trade-offs.
We find that our methods can reduce the space-time cost required by over an order of magnitude, compared to existing estimates with similar assumptions (see Fig.~\ref{fig:comparison}).

\subsection{Experimental Settings and Baseline}
Our experiments employ the architecture described above, with neutral atom physical parameters described in Sec.~\ref{sec:atom_array} and Tab.~\ref{tab:neutral_atom_params}, the circuit-level physical error model described in Sec.~\ref{sec:scaling}, and the effective logical error model described in Sec.~\ref{sec:scaling}.
Since qubits are moved around in a reconfigurable architecture, the syndrome extraction can be pipelined.
We also assume that movements for transversal $H$ and $S$ gates, which involve permutations or folds, take the same amount of time as transversal entangling gates.
We calculate the logical error of each step using Eq.~(\ref{eq:ler-per-cnot}), including all gates and errors between each pair of syndrome extraction rounds.
When transversal gate operations are preceded or followed by idle memory rounds or logical measurements, some of the errors can likely be absorbed by correction during those operations~\cite{tan2024resilience}, so our estimate is a conservative approximation.

We employ the architecture and detailed improvements described in the preceding sections, and search over parameters to minimize the space-time volume of implementing the algorithm.
We compare against highly optimized state-of-the-art compilations based on surface codes and lattice surgery~\cite{fowler2012surface,beverland2022assessing,webber2022impact,gidney2019how}, for the same 2048-bit factoring problem. 

\begin{figure}
\centering
\includegraphics[width=\linewidth]{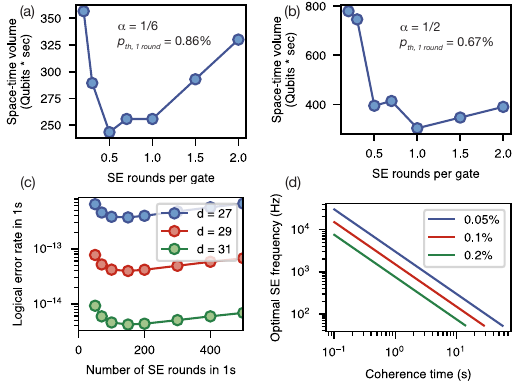}
\caption{(a,b) Optimizing the number of SE rounds in the $8T$-to-$CCZ$ factory, for two different parameters of the effective threshold under 1 QEC round per CNOT. (c,d) Optimizing the frequency of syndrome extraction in an idle block. We find that the optimal frequency is largely independent of the code distance (c), and occurs when the idle error becomes comparable with gate error contributions (d). In (d), the different lines correspond to different gate error rates.}
\label{fig:factory_fit}
\end{figure}

\begin{table}
\caption{Algorithm parameters chosen for 2048-bit factoring.}
\label{tab:factoring_params}
\centering
\begin{tabular}{|c|c|c|}
\hline
Parameter & Our work & Ref.~\cite{gidney2019how}\\\hline\hline
Exponent window $w_{\mathrm{exp}}$ & 3 & 5\\\hline
Multiplication window $w_{\mathrm{mul}}$ & 4 & 5 \\\hline
Runway separation $r_{\mathrm{sep}}$ & 96 & 1024\\\hline
Runway padding $r_{\mathrm{pad}}$ & 43 & 43\\\hline
Code distance & 27 & 27\\\hline
Max factory number & 192 & 28 \\\hline
\end{tabular}
\end{table}

\subsection{Resource Analysis for 2048-bit Shor's Factoring Algorithm}
\label{sec:cost_breakdown}

We start by examining the optimal frequency of syndrome extraction, both during gate operation and storage.
In Fig.~\ref{fig:factory_fit}(a,b), we show the space-time volume of a magic state factory under different number of SE rounds, where for each data point, we optimize the code distance in the factory to achieve a given target logical error rate.
We find that there is a slight dependency on the CNOT decoding factor $\alpha$, but around 1 SE round per gate provides a good balance.
For simplicity, we therefore assume that 1 SE round follows each transversal gate in the following estimates, though this can likely be further optimized for given gadgets.
Similarly, we examine the optimal frequency of performing syndrome extraction during idle storage in Fig.~\ref{fig:factory_fit}(c,d).
We find that the optimal frequency is largely independent of the code distance in the target regime, and roughly corresponds to when the idle error becomes comparable to the magnitude of gate errors.
For the majority of our evaluation below, we therefore assume a 10s coherence time and a QEC round for storage qubits every 8 ms.

We now proceed to optimize the algorithmic parameters of our factoring analysis.
We sweep parameters in pairs, identifying the choice of parameters that minimize the total space-time volume~\cite{gidney2019how}.
The resulting parameters are shown in Tab.~\ref{tab:factoring_params}, where we also contrast it with the parameter choices from Ref.~\cite{gidney2019how}.
Due to differences in timescales, transversal architecture, and recent improvements in magic state preparation~\cite{gidney2024magic}, the relative balance of different components and algorithm parameters are modified compared to existing compilations~\cite{gidney2019how}.
More specifically, in a transversal architecture, because Clifford operations are executed much faster, the reaction time becomes more limiting.
This favors the use of many parallel runway segments (smaller runway separation) and the use of more magic state factories to support the parallel workloads.

In Fig.~\ref{fig:comparison}, we compare our results to state-of-the-art baselines.
Our estimates directly account for the movement times, resulting in the gates in a QEC cycle taking around 400 $\mu$s.
Moving a code patch across the distance of a logical qubit takes around 500 $\mu$s, which is equal to the measurement time, thereby allowing pipelining of the ancilla measurements to happen simultaneously with the moves for transversal gates.
We further assume a 1 ms reaction time based on measurement times and expected decoding times based on recently-developed matching-based decoders for logical algorithms~\cite{cain2025fast,serra-peralta2025decoding}.
Existing estimates rely on lattice surgery operations, where pipelining ancilla measurements is not possible without increasing qubit usage, so we assume a 900 $\mu$s QEC cycle time.
We generate the estimates of Ref.~\cite{gidney2019how} for our parameters using their attachment (estimate\_costs.py) for a 900 $\mu$s QEC cycle time and different reaction times.
Ref.~\cite{beverland2022assessing} assumes 100 $\mu$s gate times and 100 $\mu$s measurement times and neglects the reaction time, but yields a larger resource estimate.
As shown in Fig.~\ref{fig:comparison}, our estimates significantly reduce the run time, by close to 50$\times$ compared to existing estimates.

\begin{figure}
\centering
\includegraphics[width=\linewidth]{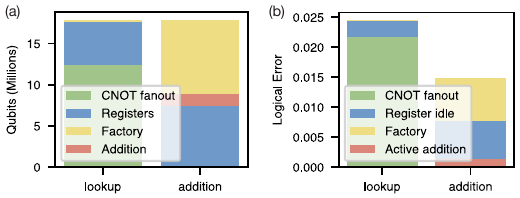}
\caption{Space usage and logical error contribution of different operations during the two main subroutines of factoring.}
\label{fig:resource_breakdown}
\end{figure}

With our tools, we can also estimate the dominant factor in the resource costs.
Our compilation results in around $1.07\times 10^6$ lookup-additions, where each lookup takes 0.17 seconds and each addition takes 0.28 seconds.
In Fig.~\ref{fig:resource_breakdown}, we show that during lookup, the CNOT fan-out dominates the space cost and logical error budget, while during addition, magic state factories dominate.

\subsection{Sensitivity Analysis}
Given the likely variability of hardware constraints and performance, we now investigate the sensitivity of our results to various parameter changes.

\subsubsection{Changes in physical error rates and decoder performance}
\label{sec:error_sensitivity}
Improvements in physical error rate alleviate demands on quantum error correction.
With lower physical error rates, one can choose to use a smaller code distance, reducing the spatial footprint.
The cost of certain resource state preparation schemes that rely on post-selection may also be substantially reduced, as the post-selection rates scale exponentially with the physical error rates~\cite{gidney2024magic}.
Furthermore, error clusters will become more localized, reducing the complexity of the decoding problem and allowing better decoders that trade speed for accuracy.

\begin{figure}
\centering
\includegraphics[width=\linewidth]{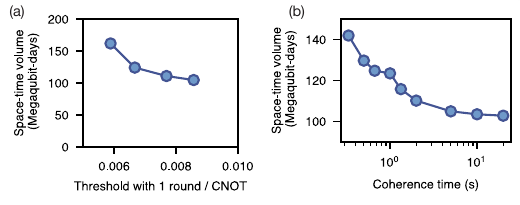}
\caption{Sensitivity of results to (a) changes in decoder performance (via varying the decoding factor $\alpha$) and (b) qubit coherence time.}
\label{fig:error_rate_sensitivity}
\end{figure}

We investigate these effects in Fig.~\ref{fig:error_rate_sensitivity}.
As we reduce the coherence time, which serves as a rough proxy for some characteristic error rates in the system, we find that the space-time volume first slowly increases, before starting to accelerate as the coherence time goes below 1s.
In Fig.~\ref{fig:error_rate_sensitivity}(a), we show how changing the CNOT decoding factor $\alpha$, which reduces the threshold under 1 SE round per CNOT, affects the space-time volume.
While reducing the threshold does increase the space-time volume, we find that the exponential scaling of errors with code distance limits the amount of increase, such that even a drop of threshold during CNOT gates from 0.86\% to 0.6\% only leads to a 50\% increase in space-time volume.

\subsubsection{Changes in timescales for physical operations and decoding}
Our resource estimations and optimizations were performed assuming timescales accessible to current hardware.
However, it is likely that future hardware improvements can significantly boost these parameters.
For example, while we assumed a 500 $\mu$s measurement time, non-destructive readout as fast as 160 $\mu$s and destructive readout as fast as 2.4 $\mu$s has been demonstrated in free space~\cite{shea2020submillisecond,ma2023high,su2024fast}, and using cavity-assisted or ensemble-assisted readout methods, readout times as short as 6 $\mu$s has been demonstrated~\cite{xu2021fast,grinkemeyer2025error,shadmany2024cavity,deist2022mid,tiecke2014nanophotonic,dordevic2021entanglement}.
For atom movement timescales, the use of higher laser powers, more tightly focused tweezers, or other innovative schemes, may also lead to faster movement times.
Finally, the use of FPGA-based or ASIC-based fast decoders~\cite{liyanage2023scalable,barber2025real}, together with improved data reuse and algorithmic techniques~\cite{zhou2024algorithmic}, can further accelerate decoding and reduce the reaction time.

\begin{figure}
\centering
\includegraphics[width=\linewidth]{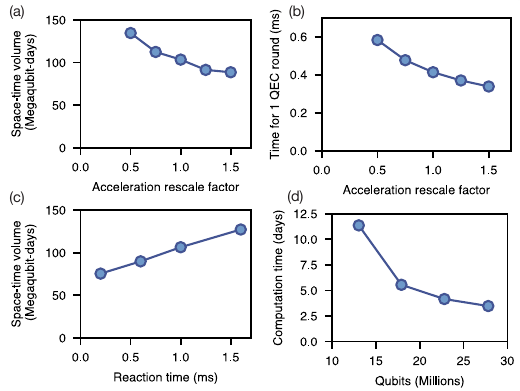}
\caption{(a) Change in space-time volume as the atom acceleration is rescaled. (b) Corresponding change in the duration of a single QEC round. (c) Change in space-time volume as the reaction time is tuned. (d) Trade-off between number of qubits employed and computation time.}
\label{fig:timescale_sensitivity}
\end{figure}

In Fig.~\ref{fig:timescale_sensitivity}(a-c), we analyze how the space-time volume changes as the timescales change.
The space-time volume improves as atom acceleration increases (Fig.~\ref{fig:timescale_sensitivity}(a)), since more operations can be performed per unit time (Fig.~\ref{fig:timescale_sensitivity}(b)).
This plot can also be used to estimate the effect of a larger or smaller grid spacing.
Another important factor is the reaction time of the control system, which we vary in Fig.~\ref{fig:timescale_sensitivity}(c).
We find that decreasing the reaction time lowers the cost, but these gains are eventually bottle-necked by the large volume of the CNOT fan-out in the lookup table.

\subsubsection{Changes in qubit number constraints}
While our analysis above has focused on optimizing the overall space-time volume, which is most relevant when considering the amortized cost across building and operating multiple systems, we also consider the case when a different trade-off is desired.
In Fig.~\ref{fig:timescale_sensitivity}(d), we evaluate the run time when qubit number is further constrained, as well as when qubit number is not a concern and we wish to further accelerate the execution.
We find that there is a continuous trade-off of parameters available with comparable total space-time volume, although the total space-time volume starts to increase more when the qubit number is below 15 million (left-most point in Fig.~\ref{fig:timescale_sensitivity}(d)).

\subsubsection{Further Optimization with Dense qLDPC Code Storage}
The flexible reconfigurability of neutral atom arrays opens the door towards incorporation of other QEC codes with much higher encoding rate, as has been proposed for various product codes~\cite{xu2024constant,tremblay2022constant} and generalized bicycle codes~\cite{bravyi2024high,viszlai2023matching}.

Due to the high encoding rate, performing selective logical gate operations on these codes are generally challenging~\cite{xu2024constant,xu2024fast,stein2024architectures,cross2024improved,cohen2022low}, although recent progress has been made towards compiling certain algorithmic gadgets onto qLDPC codes~\cite{xu2024fast}.
We therefore focus on a hybrid architecture, where logical gates are still performed with surface codes, and unused registers can be stored into a qLDPC code.
Examining the space usage in Fig.~\ref{fig:resource_breakdown}(a), we see that there are opportunities to reduce the qubit usage for storage registers, but only 4-6 million qubits are idling.
Therefore, assuming that dense qLDPC code storage can reduce the space footprint by a factor of 10~\cite{xu2024constant,tremblay2022constant,bravyi2024high,cohen2022low}, we expect $\sim$20\% decrease in the space footprint if the execution time remains the same.
We leave more careful analysis, including with improved gate constructions~\cite{cross2024improved} and the increase in QEC cycle time due to longer-distance moves for qLDPC codes~\cite{xu2024constant}, to future work.

\section{Prior Work}
\label{sec:prior_work}

\textbf{Transversal gates and fast logical operations.}
Recent experiments in neutral atom arrays and ion traps have demonstrated transversal gate operations~\cite{bluvstein2024logical,ryan-anderson2022implementing,ryan-anderson2024high,postler2022demonstration,reichardt2024logical}, providing motivation for their architectural study~\cite{shor1996fault,wang2003confinement}.
Earlier works have studied their performance, with varying assumptions about the number of required SE rounds~\cite{beverland2021cost,cai2023looped,duckering2020virtualized,dennis2002topological,viszlai2023architecture}.
More recent work has shown how to utilize correlated error information between multiple code patches to improve decoding of transversal gates and achieve $O(1)$ SE rounds per logical operation for universal quantum computation~\cite{cain2024correlated,zhou2024algorithmic,beverland2021cost,sahay2024error,wan2024iterative,cain2025fast,serra-peralta2025decoding}.
We propose new logical error models and study the compilation in a full-scale architecture, incorporating detailed layouts and full-scale resource estimation, which has not been carried out before.

It is also worth comparing the space-time savings achieved here with complementary methods based on qLDPC codes~\cite{bravyi2024high,xu2024constant,tremblay2022constant,higgott2024constructions}.
Existing analyses of resources have focused on hybrid store-compute architectures~\cite{viszlai2023matching,stein2024architectures,xu2024constant}, finding only 2$\times$ reduction in space-time volume, compared to the 50$\times$ we achieve here.

\textbf{Large-scale fault-tolerant resource estimation.}
To date, resource estimation has been primarily carried out with architectures based on lattice surgery or concatenated codes~\cite{gidney2019how,jones2012layered,fowler2012surface,ogorman2017quantum,gheorghiu2019benchmarking,babbush2018encoding,lee2021even,kim2022fault,kivlichan2020improved,campbell2021early,reiher2017elucidating,von2021quantum,litinski2022active,bellonzi2024feasibility,penuel2024feasibility,elenewski2024prospects,agrawal2024quantifying,goings2022reliably,metodi2005quantum,monroe2012large}.
Most of these focus on timescales relevant to superconducting qubits (1 $\mu$s QEC cycle time), but the estimates can be easily scaled to neutral atom timescales, typically resulting in run times on the order of several years or more.
In Sec.~\ref{sec:cost_breakdown}, we provided our methodology for adapting the analysis of Ref.~\cite{gidney2019how,beverland2022assessing} to estimate the resource, finding that existing lattice-surgery-based architectures are a factor of 50$\times$ slower than our transversal architecture.
Ref.~\cite{litinski2022active} provides methods to speed up operations, but this relies on extensive long range swaps, which requires long movement times.
Even for local gadgets, the interaction range of local gadgets was as high as 12, which would involve a movement time of 1.5 ms and result in a QEC cycle time close to 3 ms.
Scaling the estimates in that paper accordingly would then result in an estimated run time of 111 days, assuming that the long range swaps in the architecture are not used.
Refs.~\cite{metodi2005quantum,whitney2009fault} describe a transversal architecture based on concatenated codes.
However, concatenated codes result in lower thresholds and larger overheads in current experimental regimes.

To the best of our knowledge, detailed resource estimation of large-scale algorithms has not been carried out for a transversal architecture with high-threshold surface codes.
Our work shows that transversal architectures can provide substantial space-time saving in platforms that support them.

\section{Discussions and Outlook}

Our work provides a detailed analysis of a large-scale, low-overhead transversal architecture for neutral atom arrays.
Through detailed analysis of the layout, logical performance, and space-time cost, our work corroborates the substantial savings that fault-tolerant architectures based on transversal gates can provide.
Our run times represent over an order of magnitude improvement compared to prior state-of-the-art, and can likely be further improved.

There are many interesting directions that our work further motivates.
While we have focused on a monolithic neutral atom architecture, there are many hardware aspects that can be further incorporated, such as details of control, continuous reloading and readout, specialized optical tools to accelerate key bottlenecks, as well as distribution of the computation across multiple networked quantum computers~\cite{buhrman2003, monroe2012large}.
Our estimates can also be further refined through direct simulations of key subroutines, allowing further optimization of the location and frequency of syndrome extraction and evaluation of decoding times.
Depending on the parameter regime, there are also alternative algorithms~\cite{kahanamoku-meyer2024fast,gidney2020quantum} or magic state preparation methods that are worth investigating.
Finally, it will be interesting to experimentally implement scaled-down versions of some of these key subroutines, as a stepping stone towards the eventual full-scale quantum computers.

\section{Acknowledgements}
We acknowledge helpful discussions with M.~Beverland, A.~Dalzell, C.~Gidney, D.~Litinski, S.~Mcardle, J.~Thompson.
We acknowledge financial support from IARPA and the Army Research Office, under the Entangled Logical Qubits program (Cooperative Agreement Number W911NF-23-2-0219), the DARPA ONISQ program (grant number W911NF2010021), the DARPA MeasQuIT program (grant number HR0011-24-9-0359), the Center for Ultracold Atoms (a NSF Physics Frontiers Center, PHY-1734011), the National Science Foundation (grant numbers PHY-2012023 and  CCF-2313084), the NSF EAGER program (grant number CHE-2037687), the Army Research Office MURI (grant number W911NF-20-1-0082), the Army Research Office (award number W911NF2320219 and grant number W911NF-19-1-0302), the Wellcome Leap Quantum for Bio program, and QuEra Computing.
D.B. acknowledges support from the NSF Graduate Research Fellowship Program (grant DGE1745303) and The Fannie and John Hertz Foundation.
M.C. acknowledges support from Department of Energy Computational Science Graduate Fellowship under award number DE-SC0020347.

\bibliography{main}

\end{document}